# Observation of inconsistent carbon isotope compositions of chlorine-isotopologue pairs of individual organochlorines by gas chromatography-high resolution mass spectrometry

Caiming Tang[1,*], Jianhua Tan[2], Yujuan Fan[1,3], Ke Zheng[1,3], Qiuxin Huang[4], Xianzhi Peng[1]

[1] State Key Laboratory of Organic Geochemistry, Guangzhou Institute of Geochemistry, Chinese Academy of Sciences, Guangzhou 510640, China

[2] Guangzhou Quality Supervision and Testing Institute, Guangzhou, 510110, China

[3] University of Chinese Academy of Sciences, Beijing 100049, China

[4] CEPREI Environmental Assessment and Monitoring Center, The 5th Electronics Research Institute of the Ministry of Industry and Information Technology, Guangzhou 510610, China

**Running title:** Inconsistent carbon isotope compositions of chlorine-isotopologue pairs

*Corresponding author: Dr. Caiming Tang, State Key Laboratory of Organic Geochemistry, Guangzhou Institute of Geochemistry, Chinese Academy of Sciences, Guangzhou 510640, China; Tel: +86-020-85291489; Fax: +86-020-85290009; Email: CaimingTang@gig.ac.cn.

**Abbreviations: CIP**, chlorine-isotopologue pair; **EI**, electron ionization; **HCB**, hexachlorobenzene; **IR**, isotope ratio; **Me-TCS**, methyl-triclosan; **MID**, multiple ion detection; **PCB**, polychlorinated biphenyl; **PCE**, perchlorethylene; **TCE**, trichloroethylene

**Keywords:** Carbon and chlorine isotopologues; Carbon isotope ratio; Chlorine-isotopologue pair; High resolution mass spectrometry; Organochlorines

# ABSTRACT

This study investigated the consistency/inconsistency of carbon isotope compositions of chlorine-isotopologue pairs, e.g., $^{12}C_2{}^{35}Cl_4$ vs. $^{12}C^{13}C^{35}Cl_4$, of individual organochlorines including two chloroethylenes, three polychlorinated biphenyls, methyl-triclosan and hexachlorobenzene. The raw carbon isotope ratios were measured by gas chromatography-high resolution mass spectrometry. Data simulations in terms of background subtraction, background addition, dual $^{13}C$-atoms substitution, deuterium substitution and hydrogen-transfer were conducted to confirm the validity of measured carbon isotope ratios and their differences. Inconsistent carbon isotope ratios derived from chlorine-isotopologue pairs of individual organochlorines were observed, and the isotopologues of each organochlorine were thus inferred to be non-randomly distributed. Mechanistic interpretation for these findings was tentatively proposed according to a basic principle in clumped-isotope geochemistry, reaction thermodynamics and kinetics, along with isotope effects occurring on electron ionization mass spectrometry. This study sheds light on the actual carbon isotope compositions of chlorine-isotopologue pairs of organochlorines, and yields new insights into the real distributions of carbon and chlorine isotopologues. The inconsistent carbon isotope compositions of chlorine-isotopologue pairs are anticipated to benefit the exploration of formation conditions and source identification of organochlorine pollutants.

# 1 Introduction

Organochlorines, produced by both anthropogenic and natural activities [1-3], have been greatly impacting human beings and the natural environment, either positively or negatively [4,5]. Numerous organochlorines are notorious environmental pollutants, such as highly toxic polychlorinated dioxins/furans, polychlorinated biphenyls (PCBs) and dishlorodiphenyltrichloroethanes [6-8]. Conceptually, organochlorines contain at least two elements, i.e., carbon and chlorine. Both carbon and chlorine have two natural stable isotopes ($^{12}C$ vs. $^{13}C$ and $^{35}Cl$ vs. $^{37}Cl$) with certain isotope ratios in the nature, leading to characteristic isotopologue distributions of organochlorines [9]. However, little is known about the exact distributions of carbon and chlorine isotopologues of organochlorines presently.

Historically, relative abundances of carbon and chlorine isotopologues of individual organochlorines are considered to be randomly distributed, which means that the relative abundances can be calculated with the binomial theorem [10,11]. In this context, if the bulk carbon/chlorine isotope ratios of an organochlorine from different sources are identical, then the relative abundances of individual isotopologues of the organochlorine from these sources are identical. However, as reported in a large number of studies involving "clumped isotopes", the relative abundances of multiply-substituted isotopologues of some simple compounds such as carbon dioxide, oxygen, methane and nitrogen are not stochastically distributed both in principle and in practice [12-17]. It can be speculated that the chlorine isotopologues containing more than one $^{37}Cl$ atom are analogous to the multiply-substituted isotopologues in clumped-isotope geochemistry, if carbon isotopes are not taken into account. Therefore, the chlorine isotopologues may not comply with binomial distribution (stochastic distribution) theoretically. As a consequence, the carbon isotope ratio ($^{13}C/^{12}C$) derived from a chlorine-isotopologue pair (CIP), of which the two isotopologues contain 0 and 1 $^{13}C$ atom, respectively and have the same number of $^{35}Cl$ and $^{37}Cl$ atom(s) (e.g., $^{12}C_2{}^{35}Cl_4$ vs. $^{12}C^{13}C^{35}Cl_4$), may not exactly equal those of other CIPs (e.g., $^{12}C_2{}^{35}Cl_3{}^{37}Cl$ vs. $^{12}C^{13}C^{35}Cl_3{}^{37}Cl$). Yet the potentially varied carbon isotope compositions of CIPs of individual organochlorines have not been reported. In addition, revelation of the inconsistent carbon isotope compositions of CIPs may be useful to unravel the formation conditions of organochlorines and further track their sources, and thus of important significance in source identification and apportionment for organochlorine pollutants.

In this study, we used gas chromatography-high resolution mass spectrometry (GC-HRMS) to investigate carbon isotope ratios derived from CIPs of individual organochlorines including two chloroethylenes, three PCBs, methyltriclosan (Me-TCS) and hexachlorobezene (HCB). Differences among the carbon isotope ratios of CIPs were confirmed and evaluated, and mechanism explanation for the differences was proposed. This study reveals the actual carbon isotope ratios of CIPs of organochlorines, and gains new understanding towards the real distributions of carbon and chlorine isotopologues.

## 2 Materials and methods

### 2.1 Chemicals and materials

Standards of trichloroethylene (TCE, purity $\geq$ 99.0%) and perchlorethylene (PCE, purity $\geq$ 99.0%) were purchased from Tianjin Fuyu Chemical Co. Ltd. (Tianjin, China). The mixed standard solution of polychlorinated biphenyls containing PCB-18, PCB-28 and PCB-52 (10.0 µg/mL in isooctane) was bought from Accustandard Inc. (New Haven, USA). Methyl-triclosan (Me-TCS, purity: 99.5%) and hexachlorobenzene (HCB, purity: 99.5%) were bought from Dr. Ehrenstorfer (Augsburg, Germany). Full names, abbreviations, structures, and other information of the chemicals are documented in Table S1. Isooctane and n-hexane were of chromatographic grade and obtained from CNW Technologies GmbH (Düsseldorf, Germany) and Merck Corp. (Darmstadt, Germany), respectively. Reference standard perfluorotributylamine for GC-HRMS calibration was bought from Sigma-Aldrich LLC. (St. Louis, MO, USA).

The standards of TCE, PCE, Me-TCS and HCB were weighed and subsequently dissolved with n-hexane, isooctane or nonane to obtain stock solutions at 1.0 mg/mL. These stock solutions and the purchased standard solution of PCBs were further serially diluted with n-hexane, isooctane or nonane to prepare working solutions at 1.0 or 5.0 µg/mL (Table S1). All the standard solutions were kept at −20 ºC before use.

### 2.2 Instrumental analysis

The GC-HRMS system consisted of dual gas chromatographs (Trace-GC-Ultra) coupled with a double focusing magnetic-sector high resolution mass spectrometer and a Triplus auto-sampler (GC-DFS-HRMS, Thermo-Fisher Scientific, Bremen, Germany). The system control and data acquisition were performed with Xcalibur 2.0 (Thermo-Fisher Scientific). A capillary

GC column (DB-5MS, 60 m × 0.25 mm, 0.25 µm thickness, J&W Scientific, Folsom, CA, USA) was utilized, and helium was used as the carrier gas with a constant flow rate at 1.0 mL/min. The GC temperature programs are detailed in Table S1. The GC inlet and transfer line were set at 260 °C and 280 °C, respectively. The injection volume was 1.0 µL, and splitless injection mode was used.

The working parameters and conditions of the mass spectrometer are documented as follows: electron ionization (EI) source in positive mode was used; EI energy was 45 eV; ionization source was kept at 250 °C; filament current of the EI source was 0.8 mA; multiple ion detection (MID) mode was applied to data acquisition; dwell time of each isotopologue was around 20 ms; mass resolution was ≥ 10000 (5% peak-valley definition) and the MS detection accuracy was within ±0.001 u. The mass spectrometer was real-time calibrated with perfluorotributylamine during MID operation. As EI energy around 45 eV is commonly used on GC-HRMS to achieve relatively high signal intensities for molecular ions of compounds [18], therefore the EI energy at 45 eV was applied to the implementation of this study.

Chemical structures of the investigated compounds were drawn with ChemDraw (Ultra 7.0, Cambridgesoft), and the exact masses of isotopologues were calculated with mass accuracy of 0.00001 u. For isotopologues with too low relative abundances to be displayed by ChemDraw, their exact masses were manually calculated by replacing the exact relative atomic masses of the light isotopes ($^{35}Cl$ and $^{12}C$) with those of the heavy ones ($^{37}Cl$ and $^{13}C$) on the isotopologues whose relative abundances can be displayed by ChemDraw.

Only the chlorine isotopologues containing 0 or 1 $^{13}C$ atom were taken into account. For a compound with $n$ Cl atoms, all the chlorine isotopologues ($n + 1$) containing no $^{13}C$ atom were selected. In addition, all the chlorine isotopologues containing one $^{13}C$ atom were chosen except for TCE and PCE, of which only the first three were chosen (number of $^{37}Cl$ atom(s):

0-2). The mass-to-charge ratio ($m/z$) of each isotopologue ion was obtained through subtracting the mass of an electron from that of the corresponding isotopologue. The $m/z$ values were imported into the MID module for setting up monitoring method. The detailed data including retention times, isotopologue formulas, exact masses and exact $m/z$ values of the investigated compounds are provided in Table S2, and the representative chromatograms along with mass spectra are shown in Figure 1. The working solutions were directly injected onto the GC-HRMS with six analysis replicates.

## 2.3 Data processing

The comprehensive carbon isotope ratio (IR) derived from all measured isotopologues of each compound was calculated by

$$IR = \frac{\sum_{i=0}^{n} I_{bi}}{\sum_{i=0}^{n} [m \times I_{ai} + (m-1) \times I_{bi}]} \qquad (1)$$

where $n$ is the number of Cl atoms of an organochlorine; $i$ represents the number of $^{37}$Cl atoms in an isotopologue; $I_{ai}$ denotes the MS signal intensity of the isotopologue $i$ containing no $^{13}$C atom; $I_{bi}$ represents the signal intensity of the isotopologue $i$ containing one $^{13}$C atom; $m$ is the number of C atoms of the organochlorine. This calculation scheme originates from a previously reported scheme for calculating chlorine/bromine isotope ratios [19,20]. For TCE and PCE, because only the first three chlorine isotopologues containing one $^{13}$C atom were monitored, the calculation scheme of the comprehensive carbon isotope ratio was thus altered to

$$IR = \frac{\sum_{i=0}^{2} I_{bi}}{\sum_{i=0}^{2} (2I_{ai} + I_{bi})} \qquad (2)$$

The carbon isotope ratio derived from each CIP ($IR_i$) was calculated with

$$IR_i = \frac{I_{bi}}{m \times I_{ai} + (m-1) \times I_{bi}} \qquad (3)$$

The definition of CIPs is illustrated in Figure 2 with a simulated mass spectrum of a hypothesized organochlorine, and the details of CIPs of all the investigated compounds are provided in Table S2. The average MS signal intensity of each isotopologue derived from each whole chromatographic peak was used for calculating isotope ratios. Background subtraction was carried out prior to exporting MS signal intensities by subtracting baseline intensities neighboring both ends of the corresponding peak. Data from replicated measurements were applied to calculating the mean isotope ratios and standard deviations ($1\sigma$).

## 2.4 Data simulations for confirming the validity of measured isotope ratios

Theoretical abundances of isotopologues were simulated by the isotope modelling program embedded in MassLynx V4.1 (Waters Corp., Manchester, UK). The simulation was based on the binomial theorem. The simulated abundances were calibrated with the measured comprehensive carbon isotope ratios with the processes detailed in the *Supporting Information*. In this way, the simulated comprehensive carbon isotope ratios derived from the calibrated abundances of isotopologues equaled the measured. The calibrated abundances were then subjected to simulated background subtraction and addition. The simulated abundances, calibrated abundances, and calibrated abundances with simulated background subtraction/addition are provided in Table S3.

Substitution of two $^{13}$C atoms, deuterium substitution, and hydrogen-transfer (H-transfer) might influence the measured isotope ratios and thereby the observed isotope-ratio differences among CIPs. We conducted the data simulations in terms of dual $^{13}$C-atoms substitution, deuterium substitution and H-transfer to obtain simulated isotope ratios along with isotope-ratio differences for confirming the validity of measured isotope ratios.

## 2.5 Statistical analysis

Statistical analysis was carried out with SPSS Statistics 19.0 (IBM Inc., Armonk, USA). Paired-samples T test was applied to the determination of p-values (2-tailed) with alpha of 0.01 as the cut-off for significance. If a p-value is under 0.01, the null hypothesis (e.g., no difference between two groups of carbon isotope ratios) is declined, demonstrating a significant difference indeed present.

# 3 Results and discussion

## 3.1 Validity of measured carbon isotope ratios

### 3.1.1 Impact of background subtraction

To confirm the validity of measured carbon isotope ratios, we conducted some data processing procedures to examine the possible influences of instrumental uncertainties and artificial errors. We speculated that the background subtraction of MS signal intensities might negatively affect the measured carbon isotope ratios, because it might take away slight real signals of target ions. Therefore, we acquired the carbon isotope ratios with and without background subtraction, and compared them for examining the influence. In addition, we conducted data simulation for the background subtraction, in order to further confirm the validity of measured isotope ratios. The measured carbon isotope ratios with/without background subtraction and the simulated carbon isotope ratios are provided in Table S4.

As for PCB-28, Me-TCS and PCB-52, the measured isotope ratios with and without background subtraction presented increasing discrepancies from the first to the last CIPs (Figure 3a-3c), and the isotope ratios with background subtraction were significantly lower than those without background subtraction for CIP-4 of individual compounds with the discrepancies ranging from $-0.00111\pm0.00016$ to $-0.00075\pm0.00007$ (p $\leq 0.00003$, Table S5). The measured isotope ratios with background subtraction were well consistent with those without background subtraction for individual CIPs of HCB (Figure 3d). The simulated isotope ratios without background subtraction (i.e., simulated theoretical isotope ratios) of all the CIPs of each compound are consistent in theory. The simulated isotope ratios with background subtraction successively decrease from the first CIPs to the last, whereas those with background addition continually increase from the first to the last (Figure 3). The three types of simulated isotope ratios of the first CIP are very close for each compound, and close to the corresponding comprehensive isotope ratio. Whereas simulated isotope ratios with background subtraction/addition of the last CIPs of PCB-18, Me-TCS and PCB-52 and the last two CIPs of HCB showed apparently large deviations from the simulated theoretical isotope ratios. With respected to the measured isotope ratios with/without background subtraction, the discrepancies between these values and the simulated theoretical isotope ratios were larger for the last CIPs than for others of individual compounds (Figure 3). If the detected isotopologues were stochastically distributed and the background subtraction indeed caused signal loss for target ions, then the measured isotope ratios with background subtraction ought to match the simulated isotope ratios with background subtraction. On the other hand, the measured isotope ratios without background subtraction are supposed to match the simulated isotope ratios with background addition. In practice, however, the patterns of the measured carbon isotope ratios are definitely different from those of the simulated isotope ratios with background subtraction and addition (Figure 3). Therefore, the differences of measured carbon isotope ratios derived from CIPs are determined to be not caused by instrumental uncertainties and artificial errors but really existent.

*3.1.2 Influence from the substitution of two $^{13}C$ atoms, deuterium substitution and H-transfer*

In addition to background subtraction, substitution of two $^{13}C$ atoms, deuterium substitution, and H-transfer on EI-MS may affect the measured carbon isotope ratios, thus needing scrutiny and assessment. To this end, we further performed some data simulations to evaluate these potential impacts. We chose PCE as a model organochlorine to carry out the simulations of substitution of two $^{13}C$ atoms and H-transfer, in consideration of that PCE comprises merely two elements (i.e., C and Cl). In addition, the simulation of deuterium substitution was implemented with TCE, due to that TCE contains only one H atom in the formula.

As documented in Table S6, the theoretical carbon isotope ratios of the first three CIPs of PCE corrected with the substitution of two $^{13}$C atoms are 0.010997-0.011000, with the differences of 0.000001-0.000003. While the corresponding measured isotope-ratio differences were from 0.00040±0.00009 to 0.00095±0.00009 (p ≤ 0.00012), which were over two orders of magnitude higher than the simulated. This result indicates that the substitution of two $^{13}$C atoms can only negligibly affect the isotope-ratio differences. On the other hand, the mass difference between $^{13}$C$_2$$^{35}$Cl$_4$ and $^{12}$C$_2$$^{35}$Cl$_3$$^{37}$Cl is 0.00966 u, which can be separated at a MS resolution of 17172. Although the MS resolution applied in this study (around 10000) was less than 17172, partial separation between the two ions could be achieved, further reducing the impact of dual $^{13}$C-atoms substitution on the measured carbon isotope ratios.

In another study, we found the H-transfer ratio of the molecular ion of $^{13}$C$_6$-HCB was around 0.002. Based on this finding, we set the H-transfer ratio of the molecular ion of PCE at 0.01 to conduct the H-transfer simulation, which is anticipated to trigger larger impacts on carbon isotope ratios in contrast to the observed low H-transfer ratio, thus rendering the simulation results more convictive. As shown in Table S7, the differences among the corrected theoretical carbon isotope ratios of the first three CIPs range from −0.000001 to −0.0000002, which account for −3.77‰ to −0.36‰ of the measured isotope-ratio differences and are within the analysis uncertainties (standard deviations: 0.00007-0.00009). Therefore, the isotope-ratio differences caused by H-transfer are negligible and thereby cannot impact the measured isotope-ratio differences among CIPs. Furthermore, as provided in Table S7, the simulated theoretical carbon isotope ratios with H-transfer correction are 0.015866-0.015867 for the first three CIPs, which are extremely higher than the measured isotope ratios. This discrepancy demonstrates that the real H-transfer ratio was lower than 0.01, which further indicates the negligible influence of H-transfer on the observed isotope-ratio differences.

As shown in Table S8, the theoretical carbon isotope ratios corrected with deuterium substitution of the first three CIPs of TCE are 0.011506-0.011507, and the simulated isotope-ratio differences are $-0.000001$-0.000001, which are 2-3 orders of magnitude lower than corresponding measured values ($0.00022\pm0.00005$-$0.00053\pm0.00004$), indicating the ignorable effect of deuterium substitution on the observed isotope-ratio differences.

In conclusion, after the evaluation of possible impacts from several factors, the validity of the measured carbon isotope ratios was guaranteed, demonstrating that the observed variations of measured carbon isotope ratios of CIPs of each organochlorine were not artificial but real.

**3.2 Measured carbon isotope ratios**

Currently, no available study has demonstrated whether the carbon isotope ratios derived from different CIPs of individual organochlorines are exactly identical. If they are not identical, yet it is unclear whether the discrepancies can be measured by available analytical techniques. In this study, we used the scheme as expressed by Eq (3) to obtain the carbon isotope ratio of each CIP. As shown in Figure 4, Tables S-9 and S-10, the main measured carbon isotope ratios of CIPs of individual organochlorines can be differentiated, with the differences between two random CIPs several times higher than the standard deviations and the p-values less than 0.01. For instance, the carbon isotope ratios of the first three CIPs (CIP-1 to CIP-3) of PCE were $0.01150\pm0.00005$, $0.01094\pm0.00003$ and $0.01053\pm0.00008$, respectively and those for TCE were $0.01126\pm0.00004$, $0.01106\pm0.00003$ and $0.01074\pm0.00002$ (Table S9), which were significantly different from each other for individual compounds ($p \leq 0.00012$, Table S10), with the range of differences from $0.00022\pm0.00005$ to $0.00095\pm0.00009$ (Table S10). For PCBs, Me-TCS and HCB, the carbon isotope ratios of the first CIPs were significantly higher than those of others for individual compounds (Figure 4b-4d), with the discrepancies ranging from $0.00027\pm0.00005$ to $0.00219\pm0.00029$ ($p \leq 0.00136$, Table S10). The CIP-2 and CIP-3 of PCB-18, PCB-28 and Me-TCS showed insignificantly different carbon isotope ratios ($0.01903 \leq p \leq 0.12721$), and the four middle CIPs (CIP-2 to CIP-5) of HCB also presented insignificant isotope-ratio differences ($0.01592 \leq p \leq 0.92975$), except that between the CIP-2 and CIP-3 ($p = 0.00275$). The carbon isotope ratios between CIP-3 and CIP-4 of PCB-18 and PCB-28 were insignificant different ($p > 0.01$). The carbon isotope ratios of CIP-4 of PCB-18, PCB-28, Me-TCS and HCB, together with those of the last two CIPs of PCB-52 and HCB showed generally higher standard deviations (0.00015-0.00029) compared with others (Figure 4b-4d and Table S9), possibly due to the relatively low abundances of corresponding isotopologues.

Based on the measured carbon isotope ratios, we conclude that not all the measured carbon isotope ratios derived from CIPs of individual organochlorines were exactly equivalent. To the contrary, many CIPs presented significantly different carbon isotope ratios. The observed significant isotope-ratio differences support the conclusion that carbon and chlorine isotopologues of individual organochlorines are not stochastically distributed.

### 3.3 Theoretical derivation and mechanistic interpretation

In this study, we applied a basic principle in clumped-isotope geochemistry, reaction thermodynamics and kinetics, and theories relevant to isotope effects on EI-MS to the mechanistic interpretation for the observed inconsistent carbon isotope ratios of CIPs and non-randomly distributed isotopologues of organochlorines.

#### 3.3.1 Interpretation based on the basic principle in clumped-isotope geochemistry

In clumped-isotope geochemistry, a basic principle is that the relative abundances of isotopologues of individual compounds (such as carbon dioxide, nitrogen, methane, oxygen and hydrogen) do not conform to stochastic distribution except that those compounds are generated at extremely high temperatures [12-17]. The theoretical explanation regarding the nonrandom distribution of isotopologues has been detailed in a previous review [13]. In the present study, the isotopologues containing more than one heavy isotope atom ($^{37}$Cl and/or $^{13}$C) are analogous to the multiply-substituted isotopologues in clumped-isotope geochemistry. Therefore the carbon/chlorine isotopologues of organochlorines are supposed to be non-binomially distributed if the organochlorines are not produced at extremely high temperatures, resulting in inconsistent carbon isotope ratios of CIPs. The observation of inconsistent carbon isotope ratios derived from CIPs of individual organochlorines in this study is a new evidence for the principle in clumped-isotope geochemistry. In addition, this finding demonstrates that the deviations between actual and theoretical (random) distributions of carbon/chlorine isotopologues of organochlorines are measureable by GC-HRMS.

*3.3.2 Inference in light of reaction thermodynamics and kinetics, and isotope effects on EI-MS*

The observed inconsistent carbon isotope ratios of CIPs of individual organochlorines in this study may ascribe to the chlorination reactions in synthesis. As derived in the *Supporting Information*, the carbon isotope ratios of CIPs of synthesized organochlorines are deduced to be inconsistent, no matter the chlorination reactions are thermodynamically or kinetically controlled.

In this study, the investigated organochlorines were analyzed by GC-EI-HRMS. Isotope effects occurring on EI-MS can be applied to explaining the inconsistent carbon isotope ratios of CIPs of an organochlorine whose chlorine isotopologues are hypothesized to be binomially distributed. As indicated in a previous study, fragmentation on EI-MS can cause significant hydrogen isotope effects [21]. Thus, dechlorination on EI-MS are anticipated to generate carbon and chlorine isotope effects. Due to the isotope effects on EI-MS, the carbon isotope ratios of CIPs of an organochlorine measured by EI-MS cannot be consistent, even though the chlorine isotopologues are postulated to comply with binomial distribution prior to fragmentation. The related theoretical derivation is detailed in the *Supporting Information*.

## 3.5 Application prospects

As revealed in clumped-isotope geochemistry, clumping isotope effects strongly correlate with reaction temperatures [13]. When a reaction takes place at a low temperature, the ratios of multiply-substituted isotopologues tend to deviate from stochastic ratios. To the contrary, if a reaction occurs at a high temperature, the ratios of multiply-substituted isotopologues are prone to close to stochastic ratios [13]. We thus conclude that the inconsistency extents of carbon isotope ratios of CIPs are related to chlorination temperatures. The lower the chlorination temperature is, the more inconsistent the carbon isotope ratios of CIPs are. Therefore, the patterns of carbon isotope ratios of CIPs may be able to probe the temperature conditions of chlorination reactions, and further illuminate the reaction mechanisms. On the other hand, since organochlorines from different sources may be synthesized at different temperatures, the carbon isotope ratios of CIPs have promising application prospects in source identification and apportionment for chlorinated organic pollutants.

# 4 Concluding remarks

In this study, we systematically investigated whether the carbon isotope ratios derived from CIPs were consistent using seven exemplary organochlorines. The carbon isotope ratios were determined by GC-HRMS with sufficient precisions for evaluating the isotope-ratio discrepancies among CIPs. The experimental data were carefully processed, and the validity of measured carbon isotope ratios were confirmed by data simulations. Most of the measured carbon isotope ratios derived from CIPs were found to be significantly different, showing generally declining tendencies from the first to the last CIPs. The relative abundances of carbon and chlorine isotopologues of organochlorines were deduced to be non-randomly distributed, which well coincides with the principle in clumped-isotope geochemistry, reaction thermodynamics and kinetics, and isotope effects occurring on EI-MS. The carbon isotope compositions of CIPs are anticipated to be compound-specific and source-specific, and thus can be used as fingerprint features to trace sources of organochlorines in the future. The experimental methods and data processing approaches applied in this study can be extrapolated to isotopologues containing other elements such as bromine, sulfur and silicon for revealing actual isotope compositions. The results of this study provide a prospection that the isotopologues of brominated, sulphureted and silicified organic compounds are also non-stochastically distributed. Further studies is worthwhile in terms of applications of the inconsistent isotope ratios of CIPs to source identification and apportionment for organochlorine pollutants.

## Supporting information

Additional *Supporting Information* may be found in the online version of this article at the publisher's website.

## Acknowledgements

This study was financially supported by the National Natural Science Foundation of China (Grant No. 41603092) and Science & Technology Planning Project of Guangdong Province, China (Grant No. 2016A040403039).

## Conflict of interest statement

The authors have no conflict of interest to declare.

# References

[1] Ali U., Syed J. H., Malik R. N., Katsoyiannis A., Li J., Zhang G., Jones K. C., Organochlorine pesticides (OCPs) in South Asian region: a review. *Sci. Total Environ*. 2014, *476*, 705-717.

[2] Meharg, A. A., Killham, K., Environment: A pre-industrial source of dioxins and furans. *Nature* 2003, *421*, 909-910.

[3] Keppler, F., Borchers, R., Pracht, J., Rheinberger, S., Schöler, H. F., Natural formation of vinyl chloride in the terrestrial environment. *Environ. Sci. Technol.* 2002, *36*, 2479-2483.

[4] Köhler, H. R., Triebskorn, R., Wildlife ecotoxicology of pesticides: can we track effects to the population level and beyond? *Science* 2013, *341*, 759-765.

[5] Konradsen, F., Manuweera, G., Van den Berg, H., Global trends in the production and use of DDT for control of malaria and other vector-borne diseases. *Malaria J.* 2017, *16*, 401-408.

[6] Chan, J. K. Y., Wong, M. H., A review of environmental fate, body burdens, and human health risk assessment of PCDD/Fs at two typical electronic waste recycling sites in China. *Sci. Total Environ.* 2013, *463*, 1111-1123.

[7] Malisch, R., Kotz, A., Dioxins and PCBs in feed and food—review from European perspective. *Sci. Total Environ.* 2014, *491*, 2-10.

[8] Mi, X. B., Su, Y., Bao, L. J., Tao, S., Zeng, E. Y., Significance of cooking oil to bioaccessibility of dichlorodiphenyltrichloroethanes (DDTs) and polybrominated diphenyl ethers (PBDEs) in raw and cooked fish: implications for human health risk. *J. Agric. Food Chem.* 2017, *65*, 3268-3275.

[9] Pena-Abaurrea, M., Jobst, K. J., Ruffolo, R., Helm, P. A., Reiner, E. J., Identification of potential novel bioaccumulative and persistent chemicals in sediments from Ontario (Canada) using scripting approaches with GC×GC-ToF MS analysis. *Environ. Sci. Technol.* 2014, *48*, 9591-9599.

[10] Ahmed, Z., Zeeshan, S., Huber, C., Hensel, M., Schomburg, D., Münch, R., Eisenreich W., Dandekar, T., Software LS-MIDA for efficient mass isotopomer distribution analysis in metabolic modelling. *BMC Bioinformatics* 2013, *14*:218, 1-11.

[11] Anderegg, R. J., Selective reduction of mass spectral data by isotope cluster chromatography. *Anal. Chem.* 1981, *53*, 2169-2171.

[12] Ono, S., Wang, D. T., Gruen, D. S., Lollar, B. S., Zahniser, M. S., McManus, B. J., Nelson, D. D., Measurement of a doubly substituted methane isotopologue, $^{13}CH_3D$, by tunable infrared laser direct absorption spectroscopy. *Anal. Chem.* 2014, *86*, 6487-6494.

[13] Eiler, J. M., "Clumped-isotope" geochemistry—the study of naturally-occurring, multiply-substituted isotopologues. *Earth Planet. Sci. Lett.* 2007, *262*, 309-327.

[14] Wang, Z., Schauble, E. A., Eiler, J. M., Equilibrium thermodynamics of multiply substituted isotopologues of molecular gases. *Geochim. Cosmochim. Ac.* 2004, *68*, 4779-4797.

[15] Stolper, D. A., Lawson, M., Davis, C. L., Ferreira, A. A., Neto, E. S., Ellis, G. S., Lewan, M. D., Martini, A. M., Tang, Y., Schoell, M., Sessions, A. L., Eiler, J. M., Formation temperatures of thermogenic and biogenic methane. *Science* 2014, *344*, 1500-1503.

[16] Yeung, L. Y., Ash, J. L., Young, E. D., Biological signatures in clumped isotopes of $O_2$. *Science* 2015, *348*, 431-434.

[17] Yeung, L. Y., Li, S., Kohl, I. E., Haslun, J. A., Ostrom, N. E., Hu, H., Fischer T. P., Schauble E. A., Young, E. D., Extreme enrichment in atmospheric $^{15}N^{15}N$. *Sci. Adv.* 2017, *3*: eaao6741, 1-9.

[18] Tang, C., Tan, J., Quasi-targeted analysis of halogenated organic pollutants in fly ash, soil, ambient air and flue gas using gas chromatography-high resolution mass spectrometry with isotopologue distribution comparison and predicted retention time alignment. *J. Chromatogr. A* 2018, 1555, 74-88.

[19] Tang, C., Tan, J., Xiong S., Liu, J., Fan, Y., Peng X., Chlorine and bromine isotope fractionation of halogenated organic pollutants on gas chromatography columns. *J. Chromatogr. A* 2017, *1514*, 103-109.

[20] Tang, C., Tan, J., Simultaneous observation of concurrent two-dimensional carbon and chlorine/bromine isotope fractionations of halogenated organic compounds on gas chromatography. *Anal. Chim. Acta* 2018, *1039*, 172-182.

[21] Derrick, P. J., Isotope effects in fragmentation. *Mass Spectrom. Rev.* 1983, *2*, 285-298.

## Figure legends

**Figure 1.** Representative chromatograms and high resolution mass spectra of the investigated organochlorines. TCE: trichloroethylene, PCE: tetrachloroethylene, PCB: polychlorinated biphenyl, Me-TCS: methyl-triclosan, HCB: hexachlorobenzene, NL: nominal level, *m/z*: mass to charge ratio.

**Figure 2.** Simulated mass spectrum (molecular ion) of an imaginary organochlorine on electron ionization mass spectrometry and illustration of the definition of chlorine-isotopologue pairs (CIPs). The formula of the compound is postulated to be $C_mCl_n$ with the omission of other elements; Group a ($a_0$-$a_n$) corresponds to chlorine isotopologues of which all the carbon atoms are $^{12}C$; Group b ($b_0$-$b_n$) corresponds to chlorine isotopologues of which the carbon atoms contain one $^{13}C$ atom; $a_i$ ($^{12}C_m{}^{35}Cl_{n-i}{}^{37}Cl_i$) and $b_i$ ($^{12}C_{m-1}{}^{13}C\,{}^{35}Cl_{n-i}{}^{37}Cl_i$) constitute a chlorine-isotpologue pair [CIP-(i+1)]; i denotes the number of $^{37}Cl$ atom(s) of a chlorine isotopologue.

**Figure 3.** Measured carbon isotope ratios with/without background subtraction, and simulated carbon isotope ratios with/without background subtraction or with background addition. IR: isotope ratio ($^{13}C/^{12}C$); Mea_with BS: measured carbon isotope ratios with background subtraction; Mea_without BS: measured carbon isotope ratios without background subtraction; Sim_theoretical: theoretically simulated carbon isotope ratios based on the binomial theorem and a hypothesis that the theoretically simulated comprehensive carbon isotope ratio of each organochlorine equals to the measured comprehensive carbon isotope ratio of the organochlorine (the theoretically simulated carbon isotope ratios of CIPs of individual organochlorines are identical in theory); Sim_with BS: simulated carbon isotope ratios with background subtraction; Sim_with BA: simulated carbon isotope ratios with background addition; Data simulation details are provided in the *Supporting Information*; Error bars denote the standard deviations (1 σ, n = 6).

**Figure 4.** Measured carbon isotope ratios derived from the CIPs of the investigated organochlorines. Error bars represent the standard deviations.

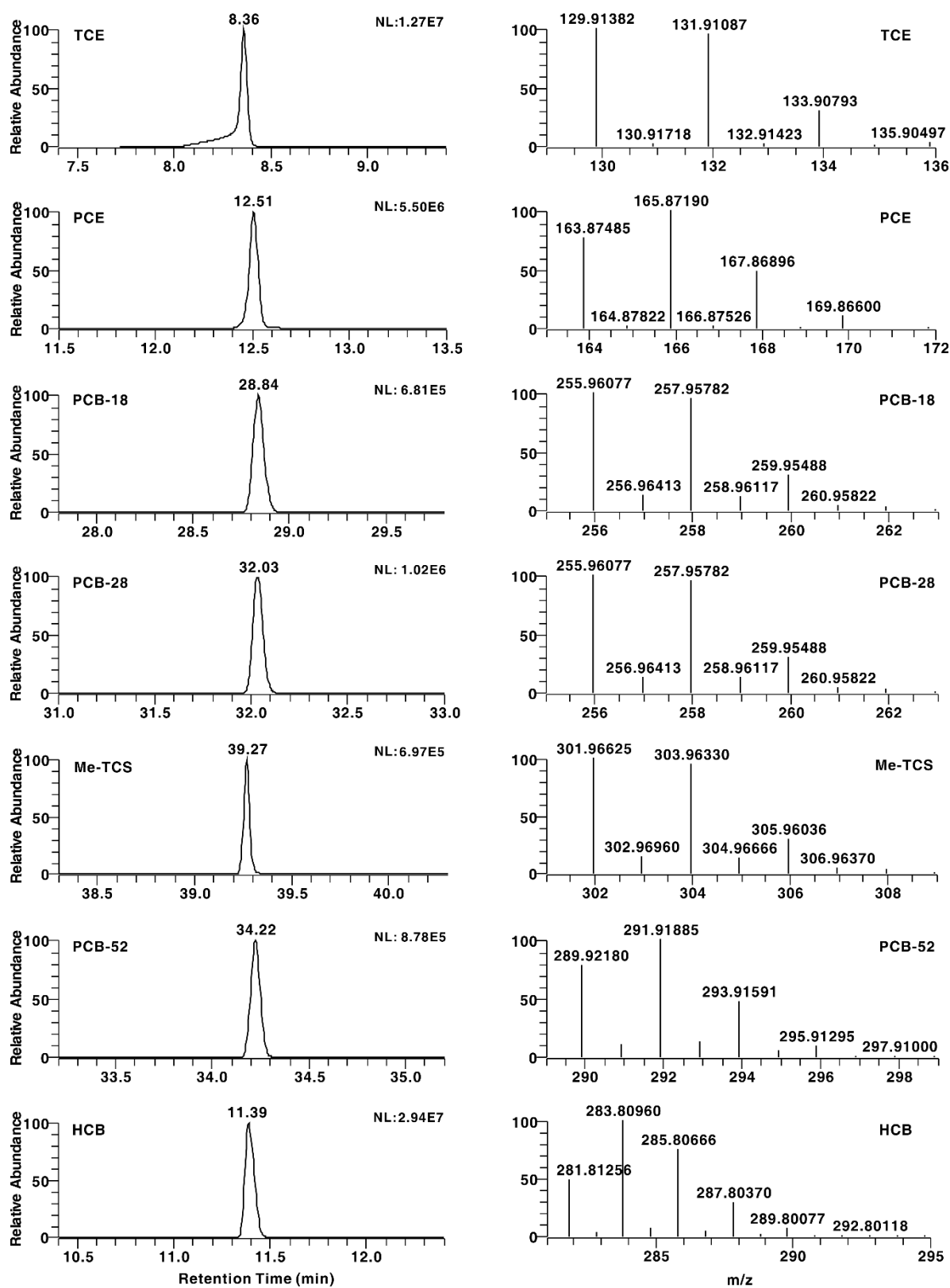

**Figure 1.** Representative chromatograms and high resolution mass spectra of the investigated organochlorines. TCE: trichloroethylene, PCE: tetrachloroethylene, PCB: polychlorinated biphenyl, Me-TCS: methyl-triclosan, HCB: hexachlorobenzene, NL: nominal level, *m/z*: mass to charge ratio.

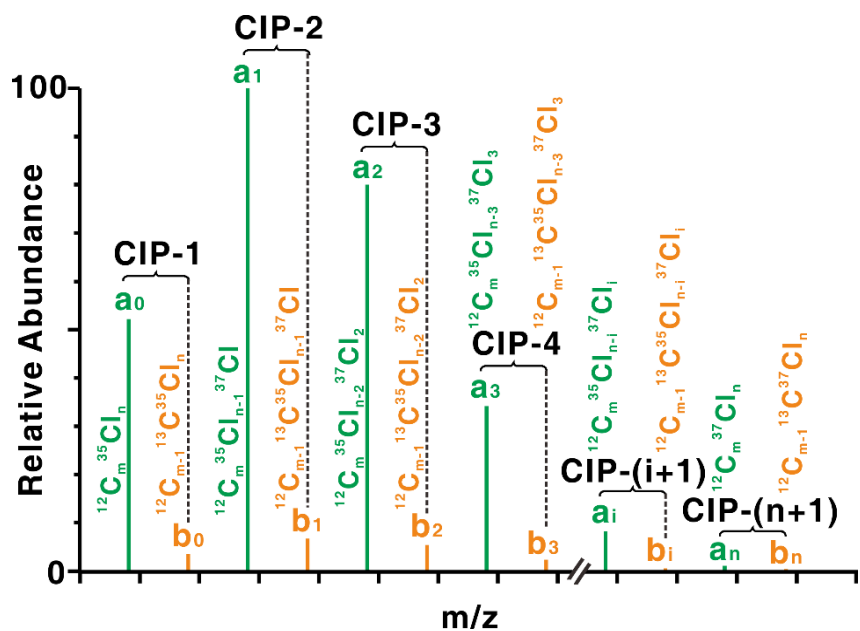

**Figure 2.** Simulated mass spectrum (molecular ion) of an imaginary organochlorine on electron ionization mass spectrometry and illustration of the definition of chlorine-isotopologue pairs (CIPs). The formula of the compound is postulated to be $C_mCl_n$ with the omission of other elements; Group a ($a_0$-$a_n$) corresponds to chlorine isotopologues of which all the carbon atoms are $^{12}C$; Group b ($b_0$-$b_n$) corresponds to chlorine isotopologues of which the carbon atoms contain one $^{13}C$ atom; $a_i$ ($^{12}C_m{}^{35}Cl_{n-i}{}^{37}Cl_i$) and $b_i$ ($^{12}C_{m-1}{}^{13}C\,^{35}Cl_{n-i}{}^{37}Cl_i$) constitute a chlorine-isotpologue pair [CIP-(i+1)]; i denotes the number of $^{37}Cl$ atom(s) of a chlorine isotopologue.

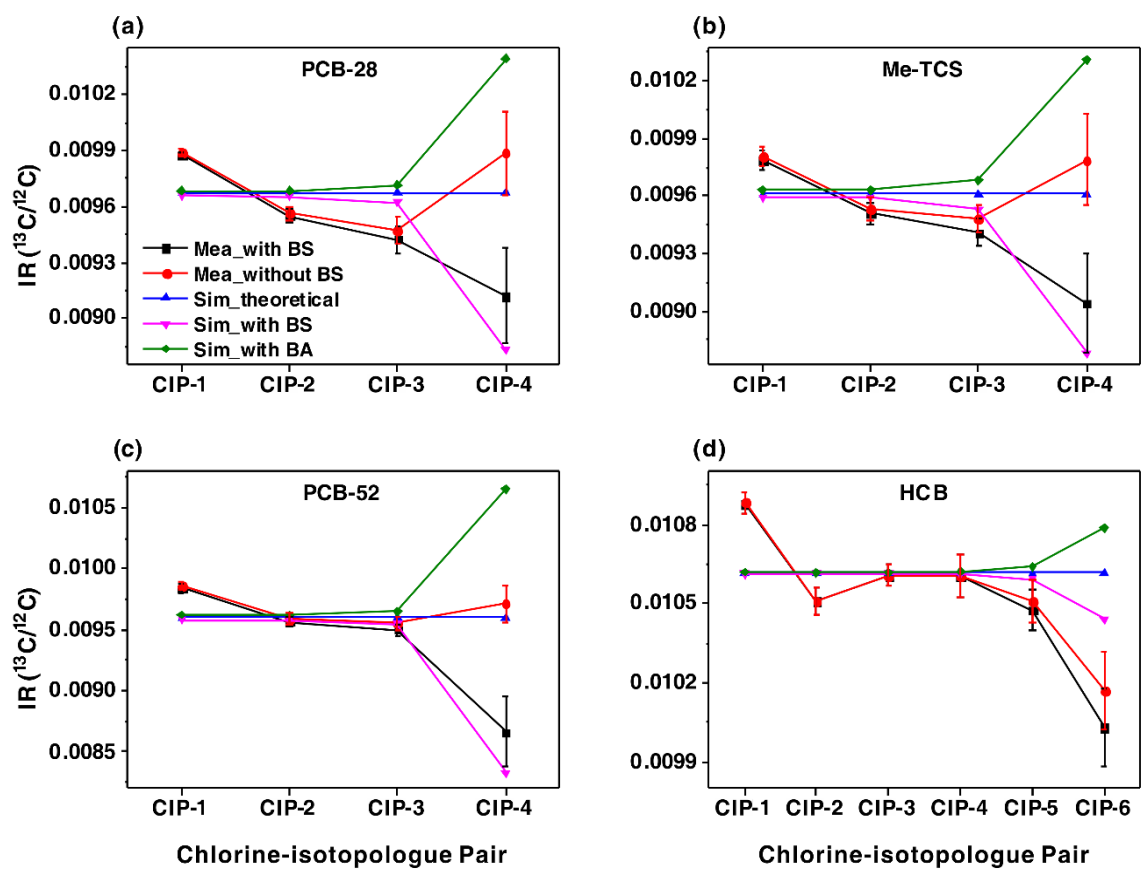

**Figure 3.** Measured carbon isotope ratios with/without background subtraction, and simulated carbon isotope ratios with/without background subtraction or with background addition. IR: isotope ratio ($^{13}C/^{12}C$); Mea_with BS: measured carbon isotope ratios with background subtraction; Mea_without BS: measured carbon isotope ratios without background subtraction; Sim_theoretical: theoretically simulated carbon isotope ratios based on the binomial theorem and a hypothesis that the theoretically simulated comprehensive carbon isotope ratio of each organochlorine equals to the measured comprehensive carbon isotope ratio of the organochlorine (the theoretically simulated carbon isotope ratios of CIPs of individual organochlorines are identical in theory); Sim_with BS: simulated carbon isotope ratios with background subtraction; Sim_with BA: simulated carbon isotope ratios with background addition; Data simulation details are provided in the *Supporting Information*; Error bars denote the standard deviations (1 σ, n = 6).

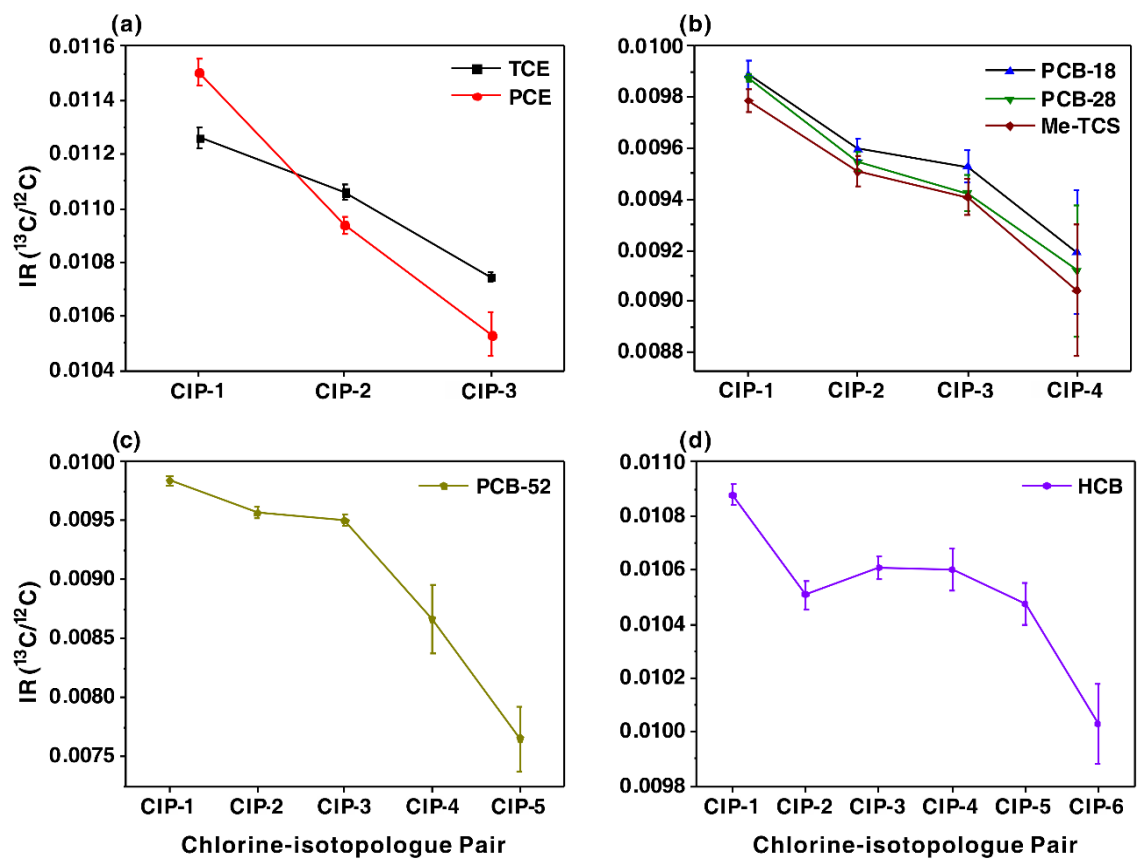

**Figure 4.** Measured carbon isotope ratios derived from the CIPs of the investigated organochlorines. Error bars represent the standard deviations.